\documentclass[twocolumn]{aastex631}

\newcommand\xmm{\textit{XMM-Newton}}
\newcommand\chandra{\textit{Chandra}}

\newcommand\xrism{\textit{XRISM}}

\newcommand\asm{\textit{RXTE/ASM}}

\newcommand\delcstat{$\Delta$C-stat}
\newcommand\pcm{cm$^{-2}$}

\newcommand\logxi{$\log (\xi$/erg~cm~s$^{-1})$}

\shorttitle{Her X-1 disk wind during Short High}
\shortauthors{Kosec et al.}
\graphicspath{{./}{figures/}}

\begin{document}

\title{Accretion disk wind of Hercules X-1 during the Short High state}

\correspondingauthor{P. Kosec}
\email{pkosec@mit.edu}

\author{P. Kosec}
\affiliation{MIT Kavli Institute for Astrophysics and Space Research, Massachusetts Institute of Technology, Cambridge, MA 02139}

\author{E. Kara}
\affiliation{MIT Kavli Institute for Astrophysics and Space Research, Massachusetts Institute of Technology, Cambridge, MA 02139}

\author{A. C. Fabian}
\affiliation{Institute of Astronomy, Madingley Road, CB3 0HA Cambridge, UK}

\author{C. Pinto}
\affiliation{INAF - IASF Palermo, Via U. La Malfa 153, I-90146 Palermo, Italy}

\author{I. Psaradaki}
\affiliation{University of Michigan, Dept. of Astronomy, 1085 S University Ave, Ann Arbor, MI 48109, USA}

\author{D. Rogantini}
\affiliation{MIT Kavli Institute for Astrophysics and Space Research, Massachusetts Institute of Technology, Cambridge, MA 02139}

\author{R. Staubert}
\affiliation{Institut für Astronomie und Astrophysik, Universität Tübingen, Sand 1, 72076 Tübingen, Germany}

\author{D. J. Walton}
\affiliation{Centre for Astrophysics Research, University of Hertfordshire, College Lane, Hatfield AL10 9AB, UK}



\begin{abstract}

Hercules X-1 is a nearly edge-on X-ray binary with a warped, precessing accretion disk, which manifests through a 35-day cycle of alternating High and Low flux states. This disk precession introduces a changing line of sight towards the X-ray source, through an ionized accretion disk wind. The sightline variation allows us to uniquely determine how the wind properties vary with height above the disk. All the previous wind measurements were made in the brighter Main High state of Her X-1. Here, we analyze the only \chandra\ observation during the fainter `Short' High state, and significantly detect blueshifted ionized absorption. We find a column density of $2.0_{-0.6}^{+1.1}\times10^{22}$ \pcm, an ionization parameter \logxi=$3.41_{-0.12}^{+0.15}$ and an outflow velocity of $380 \pm 40$ km/s. The properties of the outflow measured during the Short High state are in good agreement with those measured at equivalent precession phases during Main High. We conclude that we are sampling the same wind structure, seen during both Main and Short High, which is precessing alongside with the warped accretion disk every 35 days. Finally, the high spectral resolution of \chandra\ gratings above 1 keV in this observation enabled us to measure the abundances of certain elements in the outflow. We find Mg/O$=1.5_{-0.4}^{+0.5}$, Si/O$=1.5 \pm 0.4$ and S/O$=3.0_{-1.1}^{+1.2}$, whereas in our previous study of Her X-1 with \xmm, we found an over-abundance of N, Ne and Fe compared with O. These peculiar abundance ratios were likely introduced by pollution of the donor by the supernova which created Her X-1.

\end{abstract}



\section{Introduction} \label{sec:intro}

Hercules X-1 \citep[hereafter Her X-1,][]{Giacconi+72} is one of the most complex and fascinating X-ray binaries in the X-ray sky. Fortuitously, its proximity \citep[6.1 kpc,][]{Leahy+14} and high intrinsic luminosity ($\sim4\times10^{37}$ erg/s) permit detailed X-ray studies and allowed us to understand some of its mysteries. The compact object is a neutron star with a rotation period of 1.24 s \citep[determined from the presence of X-ray pulsations,][]{Tananbaum+72, Giacconi+73} and a surface magnetic field of $4\times 10^{12}$ G, estimated from the energy of the cyclotron resonance scattering feature \citep{Truemper+78, Staubert+19}. The neutron star is fed from an intermediate mass donor with an orbital period of 1.7 days \citep[HZ Her,][]{Davidsen+72, Bahcall+72, Middleditch+76} through a tilted, twisted and precessing accretion disk \citep{Gerend+76, Ogilvie+01}, oriented almost edge-on towards us, such that eclipses by HZ Her are observed.

The disk shape and precession manifests in the long-term evolution of Her X-1 flux, by introducing a $\sim35$ day cycle of High and Low flux states \citep{Katz+76}. The cycle begins by a turn-on into the Main High state, during which the inner accretion flow (emitting the majority of X-rays) is directly observable from our line of sight. About 10 days later, the precessing inner edge of the accretion disk starts to obscure the X-ray source. This Low State is followed by a second, fainter Short High state \citep{Fabian+73}, during which the X-ray source is just barely uncovered by the accretion disk \citep{Scott+00, Leahy+02} and Her X-1 reaches about one third of the maximum Main High state flux. Finally, a second Low State of the 35-day cycle follows, when the X-ray source is again fully covered. For an illustration of these variations, we refer to Fig. 2 of \citet{Scott+00}, which shows an average lighturve of the 35-day Her X-1 cycle, obtained by stacking many cycles observed with the \textit{RXTE} instrument (see also Fig. \ref{RXTE_lc} in this paper for an example of a single cycle).

\begin{figure*}
\begin{center}
\includegraphics[width=0.8\textwidth]{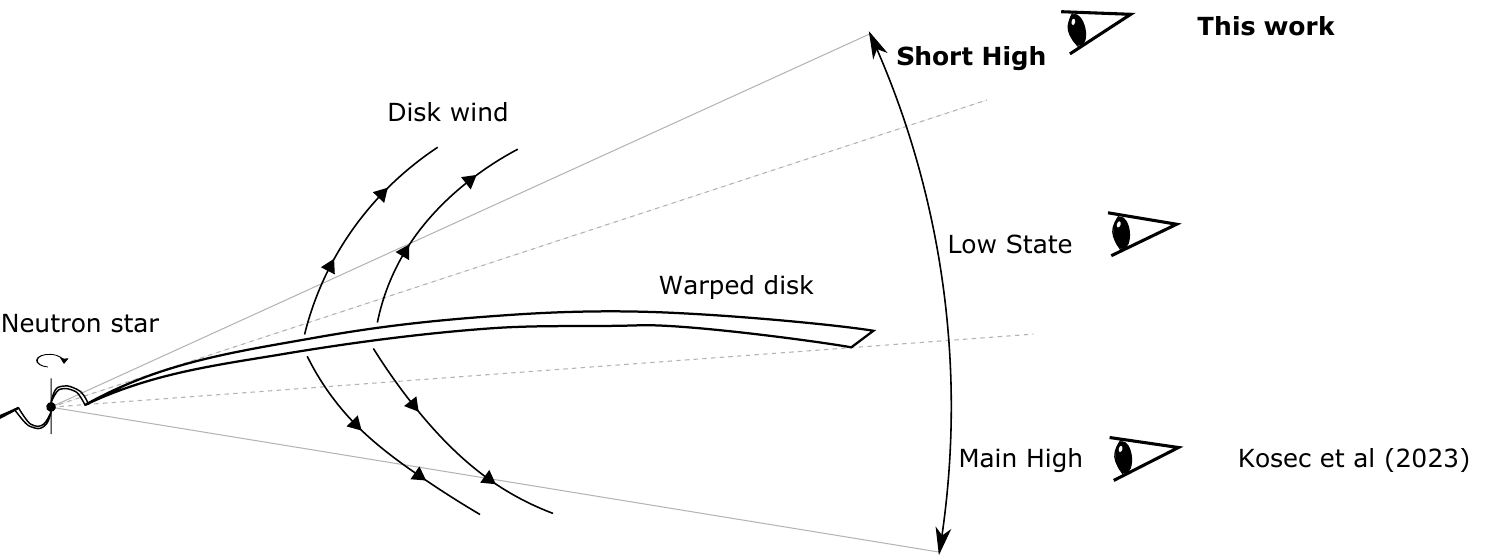}
\caption{A schematic of the Her X-1 system, adapted from Fig. 1 in \citet{Kosec+23}. In this simplified scenario, the warped precessing disk is fixed, and instead the observer line of sight is moving throughout the various states of the 35-day precession cycle. The disk precession allows our line of sight to pass through different parts of the accretion disk wind, at different heights above the disk. While \citet{Kosec+23} focused on the Main High state, this work focuses on the fainter, Short High state.} \label{HerX1_scheme}
\end{center}
\end{figure*}

The warped disk precession results in our line of sight passing through different paths through the accretion flow, and specifically at different heights above the warped disk. A unique opportunity is thus provided to us - our line of sight samples and allows us to study any ionized material over a range of heights above the disk. Namely, we are able to study the vertical distribution of the properties of the accretion disk wind of Her X-1.  A schematic of this situation is shown in Fig. \ref{HerX1_scheme}.

Accretion disk winds were detected via Doppler shifted emission or absorption lines in X-ray spectra in both black hole \citep{Kotani+00, Miller+06b, Kubota+07} and neutron star X-ray binaries \citep{Ueda+01, Miller+11} and show typical velocities of 100s km/s \citep[for a recent review, see][]{Neilsen+23}. They appear to be common especially in high inclination soft state black hole systems \citep{Ponti+12}, suggesting an equatorial wind structure. More recently, outflows were also detected in X-ray binaries outside the X-ray band - in the UV, optical and IR bands, and during both hard and soft states \citep[e.g.][]{Munoz-Darias+19, Sanchez-Sierras+20, Castro-Segura+22, Munoz-Darias+22}. 

The origin and the launching mechanisms of these phenomena are widely debated - they could be driven by Compton heating of the outer parts of the accretion disk \citep{Begelman+83, Tomaru+18}, or via magnetic fields \citep{Miller+06, Chakravorty+16, Fukumura+17}, but are unlikely to be driven by radiation pressure alone \citep{Proga+02}. These outflows have potential to drive away significant fractions of the originally accreting mass \citep[e.g.][]{Lee+02, Kosec+20} and thus affect the evolution of the X-ray binary \citep{Verbunt+93, Ziolkowski+18}. However, it is challenging to accurately constrain the wind energetics as in most systems we only observe a single line of sight towards the X-ray source. Therefore, X-ray absorption line spectroscopy gives us only a very limited view of the 3D wind structure \citep[although see][for an approach using wind re-emission]{Miller+15b}. As Her X-1 uniquely offers a range of sightlines above the accretion disk, it is an ideal source to study disk winds in order to determine their launching mechanisms and measure their energetics.

An accretion disk wind was confirmed in Her X-1 using Main High state \xmm\ observations by \citet{Kosec+20}, but previous UV observations already showed the presence of an ionized outflow within the binary system \citep{Boroson+01}. This outflow was predicted by \citet{Schandl+94} and could be the driver of the disk warping \citep[although see][for an alternative radiation-driven warping mechanism]{Ogilvie+01}. To follow-up on the opportunity to study the vertical structure of a disk wind, we performed a large \xmm\ and \chandra\ campaign, densely sampling the wind properties during a single Her X-1 Main High state \citep{Kosec+23}. We found that the wind is weaker and clumpier at greater heights above the disk, and combining the wind measurements with a Her X-1 warped disk model, we produced a 2D map of the wind properties.

Both of the X-ray studies described above were based solely on Main High state observations, where the high flux of Her X-1 results in the highest statistics spectra. However, the fact that the wind is detected in all $\sim30$ Main High state observations suggests that the wind is most likely present in Her X-1 at all times. As the outer accretion disk obscures the main X-ray source during the Low States, the detection of an ionized absorber at these periods is challenging. On the other hand, the central source is directly visible again during the Short High state, meaning it should also be possible to see the wind during this state too (Fig. \ref{HerX1_scheme}). 

Additionally, the binary system of Her X-1 and HZ Her is not oriented perfectly edge-on, but has an inclination of $\sim85^{\circ}$ \citep{Leahy+14}, which is also the average inclination of the warped accretion disk. Therefore, our line of sight during at least part of the Short High state could sample lower heights above the warped disk compared with Main High. This may happen because during the Main High, the outer disk quickly moves out of our line of sight right after the Turn-on, whereas during the Short High, different parts of the disk move away slower and remain relatively close to our line of sight for prolonged periods of time. This situation and the position of our line of sight over time with respect to different parts of the warped disk is illustrated in Fig. 2 of \citet{Scott+00} and in Fig. 3 of \citet{Leahy+02}. Following the results of the Main High state wind studies, the wind column density is expected to be higher at lower heights above the disk.

Here we study an archival high-spectral resolution \chandra\ observation of Her X-1 from Nov 2002, which reveals a number of narrow absorption lines, similar to those detected in Her X-1 spectra taken during the Main High state. The structure of this paper is as follows. Section \ref{sec:data} describes our data preparation and Section \ref{sec:modelling} the spectral modelling methods. In Section \ref{sec:results} we present the results of this study, and in Section \ref{sec:discussion} we discuss their implications. Throughout the paper, we assume a distance of 6.1 kpc to Her X-1 \citep{Leahy+14}.

\section{Data Reduction and preparation} \label{sec:data}

\chandra\ \citep{Weisskopf+02} observed Her X-1 once during its Short High state, on November 3 2002 (observation ID 4375). This is the first time this observation has been studied in detail and published. The observation exposure was 20.4 ks, and the HETG gratings \citep{Canizares+05} were used to acquire X-ray data in high spectral resolution. The midpoint of the observation was at MJD=52581.414. The data were taken in the continuous clocking (CC) mode and the mean count rate (all instruments summed) was 9.4 ct/s.

The data were downloaded from the TGCAT archive \citep{Huenemoerder+11} in fully reduced form. We use both Medium Energy Grating (MEG) and High Energy Grating (HEG) instruments, and stack the positive and negative first order spectra for each instrument using the \textsc{combine\_grating\_spectra} routine in CIAO \citep{Fruscione+06}. We analyzed the data in full spectral resolution without any binning. We also tested our final spectral fit with the optimal binning scheme \citep{Kaastra+16}, but found no differences in the final results. The HEG and MEG spectra were fitted simultaneously using a cross-calibration constant. The value of this parameter was always very close to 1, indicating $<3$\% calibration difference between the two instruments. We used HEG data in the energy range between 1 and 9 keV (1.4 to 12.4 \AA), and MEG data in the range between 0.62 keV and 5 keV (2.5 to 20\AA). These limits were imposed by the instrument calibration and the signal-to-noise.

The spectra were fitted in the \textsc{spex} fitting package version 3.06.01 \citep{Kaastra+96}. All reduced spectra were converted from \textsc{ogip} format into \textsc{spex} format using the \textsc{trafo} routine. We used the Cash statistic \citep{Cash+79}, which is appropriate for un-binned data, to analyze all spectra. The uncertainties are provided at 1$\sigma$ significance.

\section{Spectral modelling} \label{sec:modelling}

We apply the spectral model from \citet{Kosec+22} to describe the Her X-1 continuum emission during Short High. Fortunately,  we are able to make some simplifications to the complex original model due to the different instrument used here. For further details about the individual spectral components, we refer the reader to section 4 of \citet{Kosec+22}. 

A \textsc{comt} model \citep{Titarchuk+94} is used to describe the primary accretion column emission from the X-ray pulsar, and a \textsc{bb} model describes the soft (T$\sim0.1$ keV) reprocessed emission. As our coverage of the soft X-ray band ($<1$ keV) is limited, we fix the blackbody temperature to 0.1 keV. We also model Fe line emission in the Fe K band. We use two Gaussian emission lines, a narrow line at 6.4 keV (low ionization Fe), and a medium-width line at 6.7 keV (Fe XXV). A highly broadened (FWHM$\sim2$ keV) Fe K line at $\sim6.5$ keV, seen during the Main High state of Her X-1 \citep{Kosec+22}, is not required in our Short High state observation. This is likely a signal-to-noise issue - our \chandra\ HETG spectrum has much lower statistics in the Fe K band than the Main High state \xmm\ EPIC-pn spectrum. On the other hand, we do observe a strong excess at 1 keV \citep[the `1 keV feature' seen in the Main High state, e.g.][]{Fuerst+13}, which we model with a highly broadened Gaussian emission line. This excess is likely due to a forest of Fe L lines, possibly mixed with Ne IX-X emission. Finally, we also include a broadened emission line at 18.96 \AA\ (0.65 keV) to describe O VIII emission, seen also during the Main High state. The O VIII position is very close to the lower energy limit of our coverage, so we freeze the line wavelength to 18.96 \AA. As our band does not cover softer energies than 0.6 keV, we do not need to include any further lines to describe O VII, N VII and N VI emission. Finally, all of the source emission is obscured by Galactic absorption, which is described by the \textsc{hot} model as an almost neutral absorber with a temperature of $8\times10^{-6}$ keV \citep{dePlaa+04, Steenbrugge+05}. We fit for the column density of this absorber, but set a lower limit of $1 \times 10^{20}$ \pcm\ \citep{HI4PI+16}.

The absorption lines from the disk wind are described with the \textsc{pion} spectral model \citep{Miller+15, Mehdipour+16}. \textsc{pion} takes the spectral energy distribution (SED) from the currently loaded continuum spectral model and self-consistently calculates the ionizing balance and the line depths produced by an illuminated, photo-ionized slab of plasma. We use the model to determine the disk wind properties such as its column density, ionization parameter, systematic velocity, and its velocity width.

Additionally, \textsc{pion} can be used to fit for the elemental abundances in the outflow. In \citet{Kosec+23} we fitted for the abundance of N, Ne and Fe, while fixing the abundance of O to 1 (one of the elemental abundances must be fixed as all the disk wind absorption lines are from metals). Here we use the best-fitting abundances from that paper, determined from the higher statistics Main High state \xmm\ spectra: we fix N/O=3.4, Ne/O=2.3 and Fe/O=2.1. However, the broader energy band in which HETG offers very good spectral resolution (and does not suffer from instrumental issues as EPIC pn on \xmm) allows us to reliably fit for further elemental abundances: we can determine the abundance of Mg, Si and S. We have previously used 3 \chandra\ HETG spectra to study Her X-1 in the Main High state \citep{Kosec+23}, but were not able to constrain the Mg, S and Si elemental abundances because the wind was weaker in those 3 \chandra\ observations in comparison with the Short High state exposure analyzed here. The abundances of these three elements are left as free parameters in our spectral fit.

The final model, in symbolic form, is thus: \textsc{hot x pion x (comt + bb + 4 ga)}.

\section{Results} \label{sec:results}

\begin{figure}
\begin{center}
\includegraphics[width=\columnwidth]{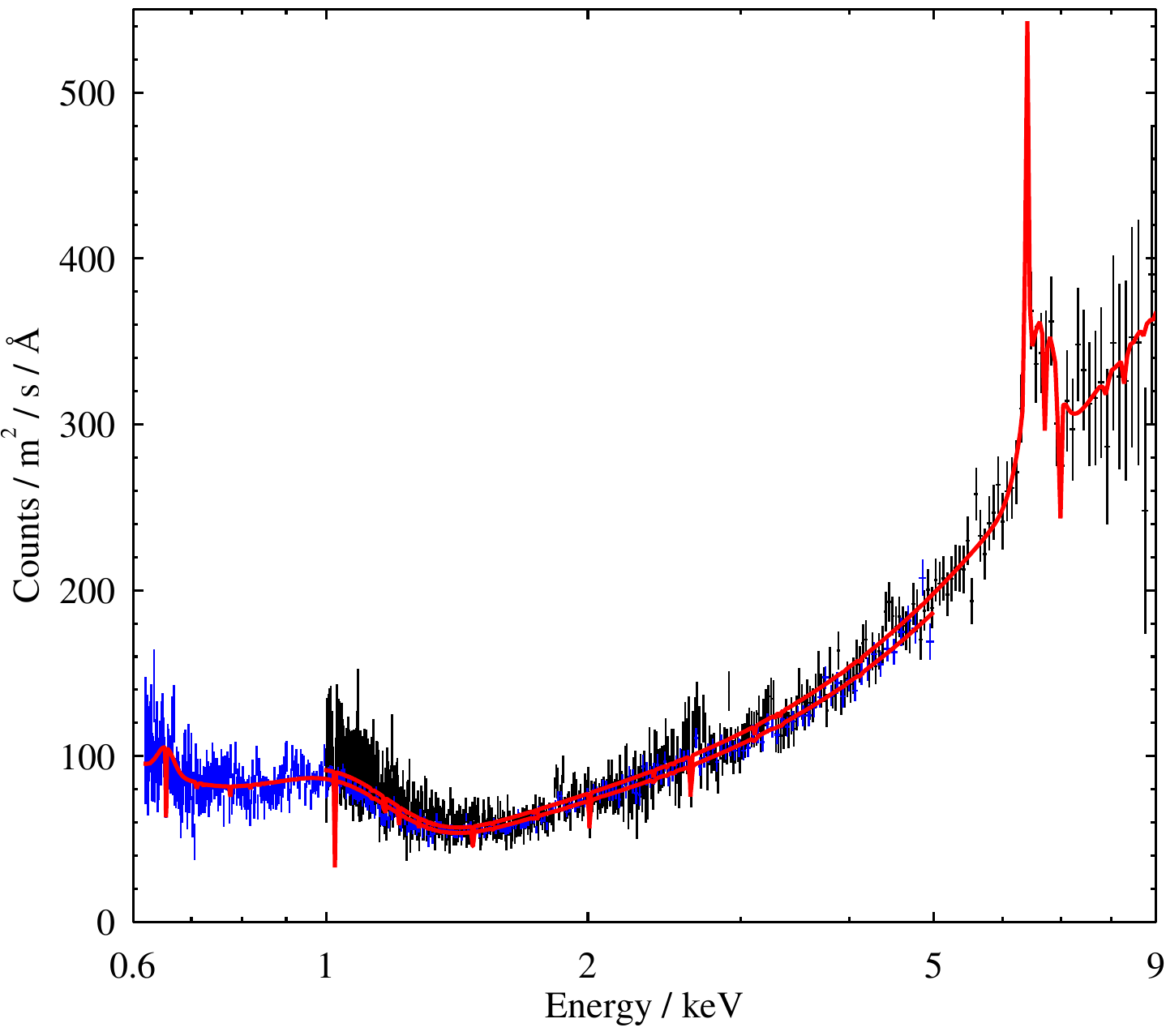}
\caption{The 0.6-9 keV Short High state \chandra\ HETG spectrum of Her X-1 (heavily overbinned), showing the broadband continuum fit with the full spectral model. MEG data are shown in blue, HEG data are in black, and the best-fitting model is in red.} \label{HETG_fullspectrum}
\end{center}
\end{figure}

\begin{figure*}
\begin{center}
\includegraphics[width=\textwidth]{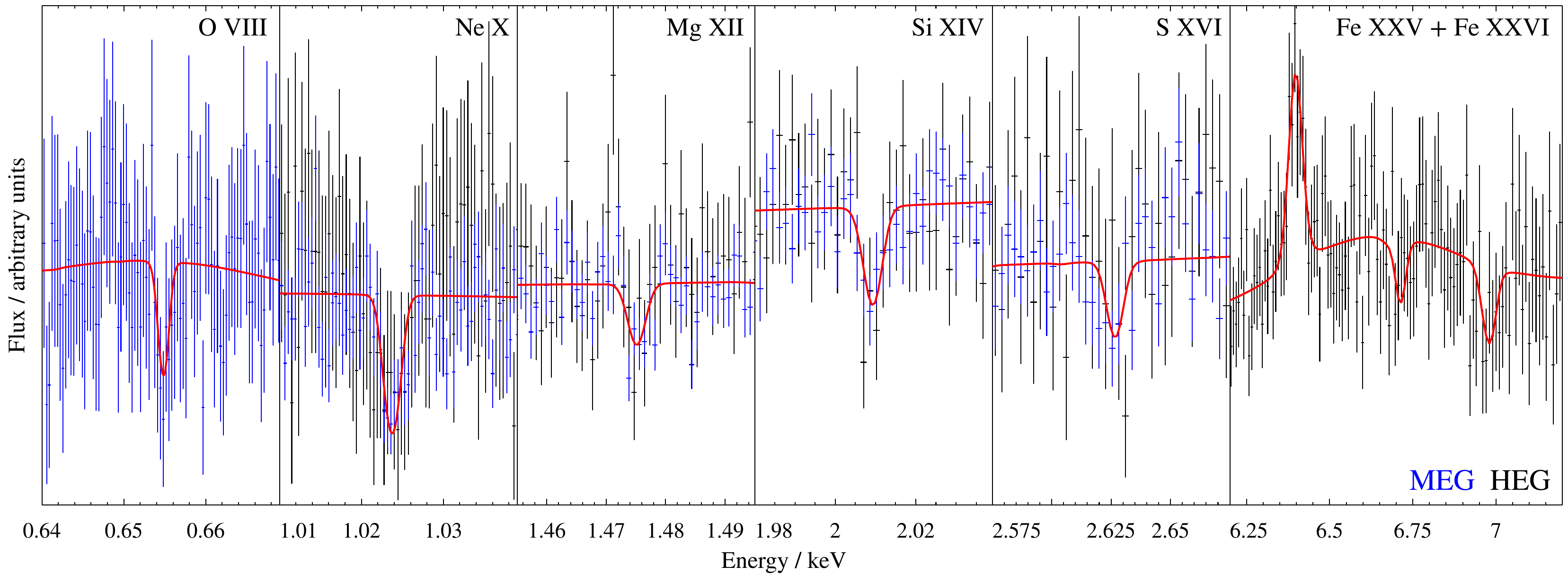}
\caption{Her X-1 spectrum from the Short High state \chandra\ observation, focusing only on the narrow energy bands of interest around the strongest elemental transitions. The absorption lines near these transitions imply the presence of an ionized disk wind. MEG data are in blue, HEG data are in black and the best-fitting spectral model (including the ionized absorption component) is in red. \label{HETG_spectrum}}
\end{center}
\end{figure*}

The high-resolution \chandra\ HETG spectrum reveals a number of absorption lines located near strong Ly$\alpha$ elemental transitions of O, Ne, Si and S. The HETG spectrum, focusing on only on the strongest elemental transitions, is shown in Fig. \ref{HETG_spectrum}. These absorption lines suggest the presence of a highly ionized outflow with a low blueshift/velocity ($<1000$ km/s). We fit the spectrum with the spectral model described in Section \ref{sec:modelling}. This results in a fit with C-stat=8599 for 7888 degrees of freedom (DoF). A continuum-only spectral fit, where the \textsc{pion} component is omitted, results in fit statistic C-stat=8767 for 7895 DoF. The very large fit statistic improvement of \delcstat=168 upon adding the ionized absorber to the baseline spectral fit, even compared to observations during the brighter Main High state \citep[Extended Data Table 1 of][]{Kosec+23}, indicates a very high statistical detection \citep[$>5\sigma$,][]{Kosec+18b} of the disk wind during this Short High state observation of Her X-1. The best-fitting spectral model is shown in Fig. \ref{HETG_fullspectrum} (which contains the full $0.6-9$ keV broadband fit) and in Fig. \ref{HETG_spectrum} (focusing only on the narrow energy bands around the strongest disk wind absorption lines).

The best-fitting disk wind properties are as follows. We determine a column density of $2.0_{-0.6}^{+1.1} \times 10^{22}$ \pcm, an ionization parameter \logxi\ of $3.41_{-0.12}^{+0.15}$, a systematic velocity of $570 \pm 40$ km/s and a velocity width $240 \pm 40$ km/s. The orbital motion-corrected outflow velocity is $380 \pm 40$ km/s, determined by following the steps from \citet{Kosec+20}. The measured elemental abundances are Mg/O=$1.5_{-0.4}^{+0.5}$, Si/O=$ 1.5 \pm 0.4$ and S/O=$3.0_{-1.1}^{+1.2}$. These results are summarized alongside all the best-fitting continuum parameters in Table \ref{results_table}. 

We tested for the multi-phase nature of the disk wind by adding a second \textsc{pion} component to the previous spectral fit. A second ionization component could potentially alter the best-fitting elemental abundances and shift them back towards Solar ratios. However, by fitting this more complex spectral model for a range of \textsc{pion} parameter values, we did not find any significant evidence (\delcstat$>9$) for a second ionization zone. Similar results were found in the Main High State \citep{Kosec+23}.

\begin{deluxetable}{ccc}
\tablecaption{Best-fitting properties of the continuum and the disk wind in the Short High spectrum of Her X-1. The continuum component luminosities are calculated from observed, absorption-corrected X-ray fluxes assuming a distance of 6.1 kpc. \label{results_table}}
\tablewidth{0pt}
\tablehead{
\colhead{Component} & \colhead{Parameter} & \colhead{Best-fitting value}  
}
\startdata 
Disk&  column density & $2.0_{-0.6}^{+1.1} \times 10^{22}$ \pcm \\
wind  & \logxi\ & $3.41_{-0.12}^{+0.15}$ \\
 & outflow velocity & $380 \pm 40$ km/s \\
 & velocity width &$240 \pm 40$ km/s\\
  & Mg/O & $1.5_{-0.4}^{+0.5}$ \\
   & Si/O & $1.5 \pm 0.4$ \\
    & S/O & $3.0_{-1.1}^{+1.2}$ \\
 & \delcstat\  & $168 $ \\
 \hline
 Primary & seed temp. & $0.33 \pm 0.02$ keV \\
 Comptonization  & electron temp. & $3.8_{-0.3 }^{+0.4}$ keV\\
& optical depth &$ 9.4 \pm 0.3$  \\
& luminosity &  $(6.5 \pm 0.4)\times 10^{36}$ erg/s\\
\hline
Soft blackbody  & temperature & 0.1 keV (fixed) \\
& luminosity &  $1.62_{-0.10}^{+0.09}\times 10^{36}$ erg/s\\
\hline
1 keV & energy & $0.91 \pm 0.02$ keV \\
feature  & FWHM &$0.46 \pm 0.02$ keV \\
& luminosity &  $(2.1 \pm 0.2)\times 10^{35}$ erg/s\\
\hline
Fe I  & energy &$6.397 \pm 0.004$ keV \\
  & FWHM & $0.037_{-0.012}^{+0.015}$ keV \\
  & luminosity &  $1.9_{-0.3}^{+0.4}\times 10^{34}$ erg/s\\
\hline
Fe XXV & energy & $6.61_{-0.04}^{+0.05}$ keV  \\
  & FWHM & $0.60_{-0.09}^{+0.13}$ keV \\
  & luminosity &  $7.4_{-1.2}^{+1.4}\times 10^{34}$ erg/s\\
\hline
O VIII & energy &  0.654 keV (fixed)\\
  & FWHM & $0.034_{-0.010}^{+0.014}$ keV\\
  & luminosity &  $8_{-3}^{+4}\times 10^{33}$ erg/s\\
\enddata
\end{deluxetable}

\section{Discussion and conclusions} \label{sec:discussion}

We study an archival Short High state observation of Her X-1 taken with the \chandra\ HETG instrument. The gratings allow us to resolve and detect the same narrow absorption lines from an ionized disk wind that are present in the Main High state observations. We model the absorber with the photo-ionization model \textsc{pion} and determine the wind properties and the abundances of certain elements within the outflow, leveraging the broader energy band of HETG in comparison with \xmm\ RGS.

\begin{figure*}
\begin{center}
\vspace{-1.3cm}
\includegraphics[width=0.8\textwidth]{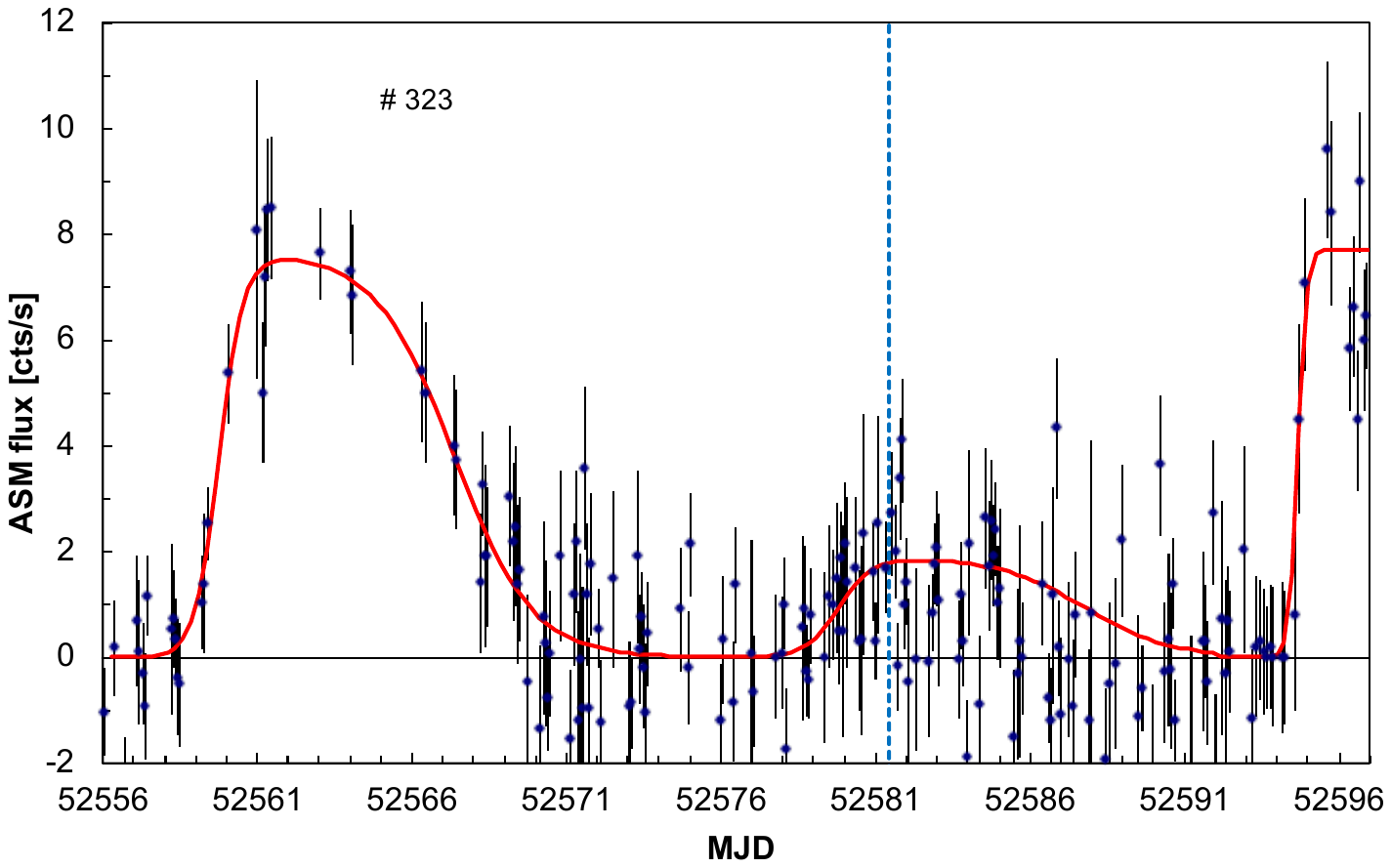}
\vspace{-1.5cm}
\caption{\asm\ lightcurve ($2-10$ keV) of Her X-1 super-orbital cycle number 323, during which the archival \chandra\ observation of the Short High state occurred. The cycle counting is according to \citet{Staubert+83}, following \citet{Davison+77}. The midpoint of the \chandra\ exposure is at MJD=52581.4 and is shown with the vertical blue dashed line. \label{RXTE_lc}}
\end{center}
\end{figure*}

To compare the properties of the disk wind during the Short High state observation with our previous study during Main High \citep{Kosec+23}, we determine the current precession phase as well as the time elapsed since the Short High state Rise for the archival \chandra\ observation. In order to estimate these quantities, we use archival RXTE/ASM observations. A lightcurve of the relevant Her X-1 precession cycle is shown in Fig. \ref{RXTE_lc}. 

The measurements of the Turn-on (into the Main High state) and the Rise (into the Short High state) are based on formal best fits with a double exponential function, which has been found to produce good fits to 35-day cycle lightcurves of Her X-1 \citep{Staubert+16}. Following the method of \citet{Staubert+16}, we determined the Main High Turn-on times for the current (MJD$=52559.3 \pm 0.1$) and the following (MJD$=52594.6 \pm 0.1$) 35-day cycles. We then estimated the precession phase of the Chandra observation to be $0.627 \pm 0.004$. To find the phase elapsed since the Short High state Rise, we also determined the Short High Rise time from the Her X-1 lightcurve to be MJD$=52579.0 \pm 0.2$. This means that the Short High Rise occurred at a precession phase of $0.559 \pm 0.009$. The mid-point of the \chandra\ observation hence happened at phase of $0.068 \pm 0.009$ since Her X-1 rose into the Short High state.

We note that the Rise into the Short High state in Her X-1 does not occur exactly one half of the 35-day cycle after the Main High Turn-on. This is because the binary system is not aligned exactly edge-on towards us, but with a more moderate inclination of about $85^{\circ}$ \citep{Leahy+14}. For this reason, for any comparisons between the Main High and the Short High state wind properties, it is necessary to correctly account for both Main High Turn-ons and Short High Rise times in order to determine the correct equivalent precession phases of both high states. Otherwise, the Short High state properties will be shifted.

\begin{figure}
\includegraphics[width=\columnwidth]{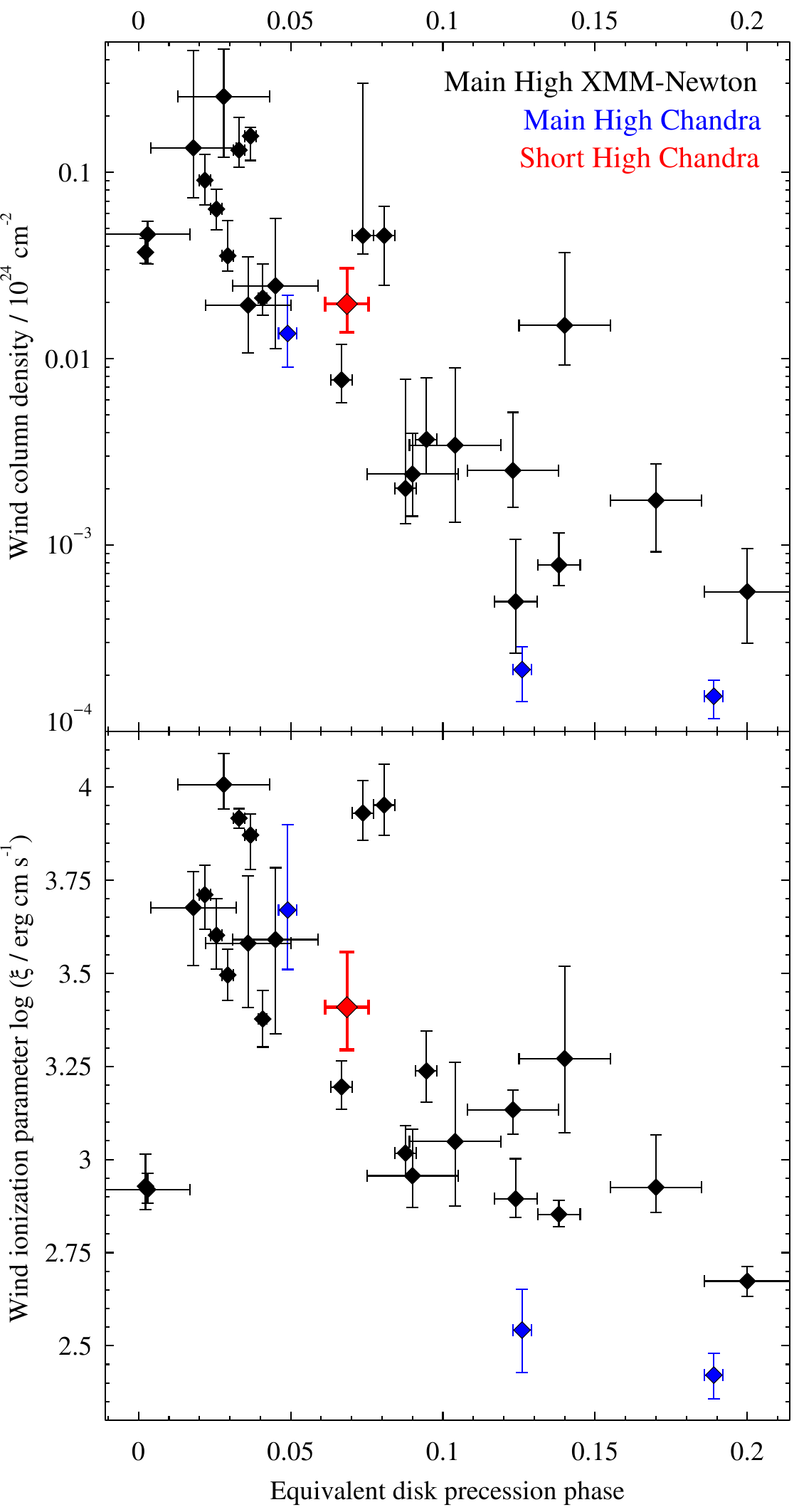}
\caption{Wind column density (top panel) and ionization parameter (bottom panel) versus the equivalent disk precession phase. Main High state observations \citep[taken from][]{Kosec+23} are shown in black (\xmm) and blue (\chandra), the only \chandra\ observation during the Short High is in red. The X-axis value of the Short High state data point is the precession phase elapsed since the Rise into the Short High state. \label{Main_Short_comparison}}
\end{figure}

In Fig. \ref{Main_Short_comparison}, we compare the best-fitting wind column density and ionization parameter during the Short High with previous measurements made during the Main High state \citep[presented in][]{Kosec+23}. We find that both the column density and the ionization parameter are in good general agreement with wind parameter measurements at equivalent precession phases during the Main High state. This suggests that we are observing the same wind structure which was detected during the Main High state, and that the structure does not vary strongly between the two high states, as it precesses alongside with the warped accretion disk.

During the Main High, we found that the wind column density decreases with increasing precession phase, i.e. with increasing height above the warped accretion disk \citep{Kosec+23}. In other words, the wind is weakening at greater heights above the disk. It is plausible that the same relationship is followed during the Short High state. This will be probed by our upcoming \xmm\ and \chandra\ observation campaign (to occur in 2023).

We can also estimate the location of the outflow during the Short High state observation and compare it with Main High measurements. Following the approach of \citet{Kosec+23}, we can determine the maximum distance of the outflow from the ionizing source such as:

\begin{equation}
    R_{\rm max} = \frac{L_{\rm{ion}}}{N_{\rm{H}}\xi}
\end{equation}

where $L_{\rm{ion}}$ is the ionizing (13.6 eV - 13.6 keV) luminosity. The estimated observed $L_{\rm{ion}}$ during the \chandra\ observation is $(6.2 \pm 0.4) \times 10^{36}$ erg/s. Therefore, the maximum distance of the outflow from the neutron star is $1.2_{-0.7}^{+0.5} \times 10^{11}$ cm. We compare this value with the maximum distances of the wind from the ionizing source during the Main High state \citep[Fig. 3 of ][]{Kosec+23}. We find that the maximum distance of the outflow from the X-ray source is roughly comparable with the maximum distances derived for wind measurements taken around phase $\mathbf{0.04-0.05}$ during the Main High state.

In addition to the column density and the ionization parameter, we also measure the wind outflow velocity and velocity width. These parameters are again consistent with those measured during Main High. The outflow velocity was observed to vary between 200 and 1000 km/s during the Main High (compared with $380 \pm 40$ km/s during Short High), while the velocity width was in most observations measured to be between 70 and 500 km/s (versus $240 \pm 40$ km/s during the Short High observation).

Thanks to the \chandra\ HETG broader bandpass (observed in high-resolution), we are additionally able to measure elemental abundances of certain elements inaccessible to \xmm. Those elements are Mg, Si and S. While technically included in the RGS band, the Mg XII line is not as well detected and resolved in the RGS data as with HETG data, preventing a direct abundance measurement. We find that both Mg and Si are about 50\% super-abundant compared to O, while the S/O ratio is as high as 3.

Combining these measurements with previous \xmm\ results \citep{Kosec+23} on other elemental abundances (N/O$=3.4_{-0.8}^{+0.6}$, Ne/O$=2.3_{-0.5}^{+0.4}$, Fe/O$=2.1 \pm 0.3$), we find that all the elements are super-abundant compared with O. This is seen at low statistical significance for Mg and Si, but is significant for the remaining 4 elements. These peculiar abundances are presumed to have originated by pollution of the donor by the supernova which created Her X-1 \citep{Jimenez+02, Jimenez+05}. An alternative explanation may be that O is depleted rather than other elements being over-abundant. However, it is not likely that O would be strongly under-abundant since the CNO cycle (presumably operating in the Her X-1 progenitor) will create a strong over-abundance of N and a strong under-abundance of C, without heavily depleting the O abundance \citep{Przybilla+10}.  The UV spectra of Her X-1 indeed show over-abundance of N and depletion of C \citep{Boroson+97}. On the other hand, how the heavier elements such as S and Fe become super-abundant by a factor of 2-3, is unclear. We note that \citet{Allen+18} found evidence for non-Solar elemental ratios of certain heavier elements including Mg and Ca in the neutron star X-ray binary GX 13+1. Similarly, \citet{Kallman+09} found an over-abundance of Fe-group elements in the black hole X-ray binary GRO J1655-40 and suggested that these findings are consistent with enrichment by a core-collapse supernova. In any case, further spectroscopic observations (of Her X-1 and other X-ray binaries) are necessary to decrease the currently significant uncertainties on the measured abundances to confirm these findings.

Finally, we also measure the properties of the various prominent emission lines of Her X-1 during Short High. The position of the narrow Fe line is fully consistent with 6.4 keV, i.e. with low ionization Fe (Fe I up to Fe XIII), while the position of the medium-width Fe line is consistent at $1\sigma$ with Fe XXV. These line energies as well as the line widths are consistent with the lines detected in Her X-1 during Main High \citep{Kosec+22}. The same is found for the width of the O VIII line, and for both the energy and the width of the 1 keV broad residual.

\begin{figure*}
\begin{center}
\includegraphics[width=0.75\textwidth]{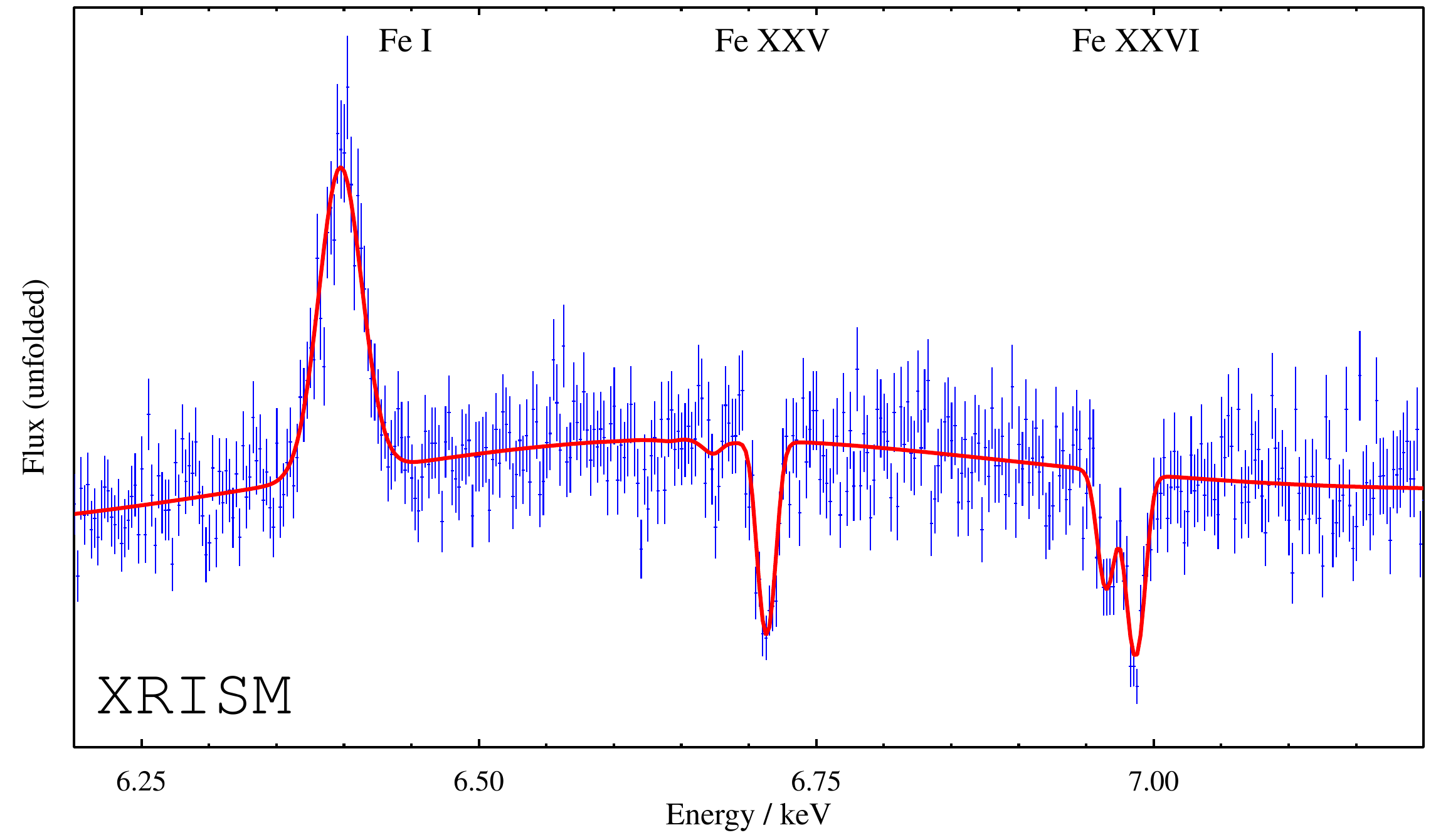}
\caption{Simulated 10 ks \xrism\ spectrum of Her X-1 in the Short High state, focusing only on the Fe K energy band. \xrism\ data are in blue and the simulated spectral model (including the ionized absorption component) is in red. The data are binned to oversample the instrumental resolution by a factor of 2. \label{XRISM_simulation}}
\end{center}
\end{figure*}

The upcoming \xrism\ observatory \citep{XRISM+20}, expected to be launched in 2023 will revolutionize the X-ray spectroscopic studies of X-ray binaries thanks to its \textit{Resolve} calorimeter with a resolution of just 5-7 eV \citep[e.g.][]{Tomaru+20}. To investigate how \xrism\ will be able to improve our measurements of the disk wind in Her X-1, we simulated a brief 10 ks observation of the Short High state using the best-fitting spectral parameters from the \chandra\ observation used in this work. We used publicly available \xrism\ simulation files\footnote{https://heasarc.gsfc.nasa.gov/docs/xrism/proposals/} assuming the goal 5 eV spectral resolution. The simulated data are shown in Fig. \ref{XRISM_simulation}, focusing just on the Fe K energy band where \xrism\ will achieve the best results. 

The \xrism\ statistics are much higher than those of \chandra\ HETG across the full energy range, in agreement with its superior effective area over \chandra. The simulated spectrum has a count rate of about 45 ct/s. At soft energies (e.g. the Ne X transition at $\sim1$ keV), the spectral resolution of \xrism\ is lower than that of \chandra\ HETG. The resolutions of the instruments are comparable at medium energies (e.g. around the S XVI transition at 2.6 keV). However, \xrism\ shines at high energies, particularly in the Fe K band, where it easily resolves individual wind absorption lines, their widths and their shapes. We particularly note that the broadening of the Fe XXV and XXVI absorption lines shown in Fig. \ref{XRISM_simulation} is not instrumental but due to the velocity width of the disk wind measured from the \chandra\ spectral fit.

We performed a spectral fit of this 10 ks \xrism\ mock simulation. The disk wind is detected over the baseline continuum with a fit improvement of \delcstat=507, indicating a much stronger detection than in the longer 20 ks archival \chandra\ dataset (where \delcstat=168 was measured). This indicates that we will be able to detect the wind during Short High and measure its properties in even shorter \xrism\ snapshots, with exposures possibly as low as 2-3 ks. Such low exposure requirement for measurements will allow us to search for any short-term wind variability. 

The wind parameters are also significantly better constrained in the 10 ks \xrism\ exposure in comparison with \chandra\ data: the column density will be determined with a precision of $2\times10^{21}$ cm$^{-2}$ (10\% precision), the ionization parameter \logxi\ will be measured with a precision of 0.04, and the outflow velocity and velocity width with a precision of 20 km/s. The uncertainties on the abundance of Mg and Si will shrink by roughly a third, while the uncertainties on the S abundance will decrease by a factor of 2. Clearly, even very brief \xrism\ snapshots will allow detailed measurements of the accretion disk wind in Her X-1 and in comparably bright X-ray binaries.

In conclusion, the \chandra\ observation of Her X-1 during the Short High state reveals the same outflow detected during the Main High. The best-fitting wind properties during Short High are broadly consistent with those determined at an equivalent precession phase during the Main High, after taking into account the delayed Short High state Rise time. By combining the results from both the Main High and the Short High, we will be able to probe the vertical structure of the disk wind of Her X-1 in great detail over a broad range of heights. This will allow us to understand, for the first time, the 3D structure and energetics of such an outflow, inaccessible in other systems with fixed sightlines without full 3D modelling of the wind re-emission. Future observations with \xrism\ (to be launched in 2023) will allow \textit{precision} measurements of the Her X-1 wind properties, particularly thanks to its superior resolution in the Fe K band. Her X-1 may thus be the Rosetta stone of X-ray binary accretion disk winds, which will allow us to fully determine their physics, energetics and 3D geometry.



\begin{acknowledgments}
 This work is based on data, obtained by the Chandra X-ray Observatory, available at \dataset[DOI: cdc.155]{https://doi.org/10.25574/cdc.155}. Support for this work was provided by the National Aeronautics and Space Administration through the Smithsonian Astrophysical Observatory (SAO) contract SV3-73016 to MIT for Support of the Chandra X-Ray Center and Science Instruments. PK and EK acknowledge support from NASA grants 80NSSC21K0872 and DD0-21125X.
\end{acknowledgments}

%

\vspace{5mm}
\facilities{\chandra
}


\software{SPEX \citep{Kaastra+96}, Veusz
          }






\bibliography{References}{}
\bibliographystyle{aasjournal}



\end{document}